\documentstyle[12pt,epsfig]{article}
\catcode`\@=11 \evensidemargin 0.0in \oddsidemargin 0.0in
\catcode`\@=11

\def\section{\@startsection{section}{1}{\z@}{3.5ex plus 1ex minus
   .2ex}{2.3ex plus .2ex}{\large\bf}}

\def\ps@headings{\def\@oddfoot{}\def\@evenfoot{}
\def\@oddhead{\hbox{}\hfill
        \makebox[.5\textwidth]{\raggedright\ignorespaces --\thepage{}--
        \hfill }}
\def\@evenhead{\@oddhead}
\def\subsectionmark##1{\markboth{##1}{}}
}

\ps@headings

\catcode`\@=12 \catcode`\@=12

\relax

\def\figcap{\section*{Figure Captions\markboth
        {FIGURECAPTIONS}{FIGURECAPTIONS}}\list
        {Fig. \arabic{enumi}:\hfill}{\settowidth\labelwidth{Fig. 999:}
        \leftmargin\labelwidth
        \advance\leftmargin\labelsep\usecounter{enumi}}}
 \relax
\def\tablecap{\section*{Table Captions\markboth
        {TABLECAPTIONS}{TABLECAPTIONS}}\list
        {Table \arabic{enumi}:\hfill}{\settowidth\labelwidth{Table 999:}
        \leftmargin\labelwidth
        \advance\leftmargin\labelsep\usecounter{enumi}}}
 \relax
\def\reflist{\section*{References\markboth
        {REFLIST}{REFLIST}}\list
        {[\arabic{enumi}]\hfill}{\settowidth\labelwidth{[999]}
        \leftmargin\labelwidth
        \advance\leftmargin\labelsep\usecounter{enumi}}}
 \relax

\catcode`\@=11

\def\marginnote#1{}
\newcount\hour
\newcount\minute
\newtoks\amorpm
\hour=\time\divide\hour by60 \minute=\time{\multiply\hour by60
\global\advance\minute by- \hour}
\edef\standardtime{{\ifnum\hour<12 \global\amorpm={am}%
    \else\global\amorpm={pm}\advance\hour by-12 \fi
    \ifnum\hour=0 \hour=12 \fi
    \number\hour:\ifnum\minute<100\fi\number\minute\the\amorpm}}
\edef\militarytime{\number\hour:\ifnum\minute<100\fi\number\minute}
\def\draftlabel#1{{\@bsphack\if@filesw {\let\thepage\relax
  \xdef\@gtempa{\write\@auxout{\string
    \newlabel{#1}{{\@currentlabel}{\thepage}}}}}\@gtempa
    \if@nobreak \ifvmode\nobreak\fi\fi\fi\@esphack}
     \gdef\@eqnlabel{#1}}
\def\@eqnlabel{}
\def\@vacuum{}
\def\draftmarginnote#1{\marginpar{\raggedright\scriptsize\tt#1}}
\def\draft{\oddsidemargin -.5truein
        \def\@oddfoot{\sl preliminary draft \hfil
        \rm\thepage\hfil\sl\today\quad\militarytime}
        \let\@evenfoot\@oddfoot \overfullrule 3pt
        \let\label=\draftlabel
        \let\marginnote=\draftmarginnote
\def\@eqnnum{(\theequation)\rlap{\kern\marginparsep\tt\@eqnlabel}%
\global\let\@eqnlabel\@vacuum}  }
\def\preprint{\twocolumn\sloppy\flushbottom\parindent 1em
        \leftmargini 2em\leftmarginv .5em\leftmarginvi .5em
        \oddsidemargin -.5in    \evensidemargin -.5in
        \columnsep 15mm \footheight 0pt
        \textwidth 250mmin      \topmargin  -.4in
        \headheight 12pt \topskip .4in
        \textheight 175mm
        \footskip 0pt

\def\@oddhead{\thepage\hfil\addtocounter{page}{1}\thepage}
        \let\@evenhead\@oddhead \def\@oddfoot{} \def\@evenfoot{}
}
\def\titlepage{\@restonecolfalse\if@twocolumn\@restonecoltrue\onecolumn
     \else \newpage \fi \thispagestyle{empty}\c@page\z@
        \def\thefootnote{\fnsymbol{footnote}} }
\def\endtitlepage{\if@restonecol\twocolumn \else  \fi
        \def\thefootnote{\arabic{footnote}}
        \setcounter{footnote}{0}}  
\catcode`@=12 \relax

\def\ps@headings{\def\@oddfoot{}\def\@evenfoot{}
\def\@oddhead{\hbox{}\hfill
        \makebox[.5\textwidth]{\raggedright\ignorespaces --\thepage{}--
        \hfill }}
\def\@evenhead{\@oddhead}
\def\subsectionmark##1{\markboth{##1}{}}
}

\ps@headings

\relax

\def\firstpage#1#2#3#4#5#6{
\begin{document}
\input epsf.tex
\newcommand{\newc}{\newcommand}
\newc{\ra}{\rightarrow}
\newc{\lra}{\leftrightarrow}
\newc{\beq}{\begin{equation}}
\newc{\eeq}{\end{equation}}
\newc{\ba}{\begin{eqnarray}}
\newc{\ea}{\end{eqnarray}}
\begin{titlepage}
\nopagebreak
\title{\begin{flushright}
        \vspace*{-0.8in}
{\normalsize hep-ph/9906238}\\[-9mm]
\end{flushright}
\vfill
{#3}}
\author{\large #4 \\[1.0cm] #5}
\maketitle \vskip -7mm \nopagebreak
\begin{abstract}
{\noindent #6}
\end{abstract}
\vfill
\begin{flushleft}
{\normalsize IOA-TH/99-6}\\[-9mm] \vspace*{.6cm} {\normalsize June
1999}\\[-9mm]
\end{flushleft}
\thispagestyle{empty}
\end{titlepage}}

\def\simlt{\stackrel{<}{{}_\sim}}
\def\simgt{\stackrel{>}{{}_\sim}}
\date{}
\firstpage{3118}{IC/95/34} {\large\bf Low Scale Unification,
Newton's Law and Extra Dimensions.}
{E.G. Floratos$^{\,a,b}$ and G.K. Leontaris$^{\,c}$}
{\normalsize\sl $^a$Institute of Nuclear Physics,  NRCS
Demokritos, {} Athens, Greece\\[-3mm] \normalsize\sl $^b$ Physics
Department, University of Iraklion, Crete, Greece\\[-3mm]
\normalsize\sl $^c$Theoretical Physics Division, Ioannina
University, GR-45110 Ioannina, Greece\\[-3mm] \normalsize\sl
} {Motivated by recent work on low energy unification, in this
short note we derive corrections on Newton's inverse square law
due to the existence of extra decompactified dimensions. In the
four-dimensional macroscopic limit we find  that the corrections
are of Yukawa type. Inside the compactified space of $n$-extra
dimensions the sub-leading term is proportional to the ($n+1$)-
power of the distance over the compactification radius  ratio.
Some physical implications of these modifications are  briefly
discussed. }
\newpage


One of the most tantalizing mysteries in modern unified theories
is the magnitude of the unification scale. A well known result in
the weakly coupled heterotic string theory is that the string
scale, is of the order of the Planck mass $M_P$~\cite{StSc}.
Recent developments have revealed the possibility that the string
scale can be arbitrarily low in Type I and Type IIB
theories~\cite{LSC,s1,JL,i1,i2,i3,li,AP,KM,LT}.

According to a recently proposed scenario, the hierarchy problem
may be solved\cite{s1}  assuming the existence of extra spatial
dimensions at low energies\cite{ia1}. In this picture, strong
gravitational effects --which could not be described accurately by
Newton's law-- may appear at short distances of the order of the
compactification scale of the extra dimensions. If so, gravitons
may propagate freely inside the space of extra dimensions, while
all ordinary particles would leave in the four dimensional world.
Experimental searches for possible deviations from Newton's
inverse square law imply that such effects should be limited below
the sub-millimeter range\cite{lcp}. We note that this scenario can
find a realization in the context of D--branes\cite{pol,jp}.
Matter fields may live in a $9$ or $3$--brane, while gravitons can
live in a larger dimensional bulk.

Deviations from the gravitational law have been intensively
studied also  in the past. In \cite{JC} the theoretical aspects of
a gravitationally repulsive term in supergravity theories were
investigated, while in \cite{TV} string loop corrections which
affect gravitational couplings were considered.

In this letter we examine corrections to the gravitational force
which are of particular importance in the case of experimental
searches  in the vicinity of the compactification radii. In the
presence of $n$ compact spatial dimensions of radii $R_{1,\dots
n}$, the fundamental scale $M_{X}$ of the theory   for very short
or very large distances can be estimated using the Gauss law. The
approximate forms of the gravitational potential in two limiting
cases in the presence of $n$-compactified extra dimensions are
given as follows\cite{s1}. Assuming for simplicity that all
compactification radii are the same $R_i=R$, inside the volume of
the extra dimensions i.e. when $r\ll R$, the Gauss low gives
 \ba
  V(r)&\sim&\frac
1{M_{P_{n+4}}^{n+2}}\frac{1}{r^{n+1}}\,, \;\;r\ll~R
\label{ap1}
\ea
where $M_{P_{n+4}}^{n+2}$ is the Planck mass in $n+4$--dimensional
space and is identified with the fundamental scale $M_X$.
 For large distances compared to the mean compactification
radius of the extra dimensions i.e., when $r\gg R$, the
$n$-dimensional compactified volume confines the gravitational
flux and as a result the approximate potential is given by
\ba
V(r)&\sim&\frac 1{M_{P_{n+4}}^{n+2}}\frac{1}{R^{n}r}\,,
\;\;r\gg~R
\ea
The latter should be identical to the known $4-d$ gravitational
potential
\ba
V(r)&=&\frac 1{M_{P_4}^2}\frac 1r
\label{ap2}
\ea
The comparison of the last two formulae for distances far beyond 
the compactification scale $M_C\sim \frac 1R$ gives an
approximate relation between the latter and the Planck mass in
$4$ and $n+4$ dimensions
\ba
\frac 1{M_C}\sim R \sim\frac{1}{M_{P_{n+4}}}
\left(\frac{M_{P_4}}{M_{P_{n+4}}}\right)^{2/n}
\ea

For distances comparable to the compactification scale corrections
are expected to modify the above formulae. In what follows, we
will present some analytic results for the case of $n=1$ and $n=2$
compactified dimensions. 
We will see that some important modifications of the above
formulae will show up in both cases. In particular, inside
the compactification circle, i.e., $r<R$,  the first sub-dominant
term will be shown to have a power dependence on the ratio $r/R$,
while at large distances the potential has a Yukawa type correction,
proportional to the form $e^{-r/R}/r$.

We will solve the Laplace equation in  $(n+3)$ spatial dimensions where $n$
of them are compactified on a torus with radius $R$. Assume the
coordinates $x_{1,2,3}$ for the 3--dimensional ordinary space and
$x_i^c,\, i=1,\cdots n$ for the compactified ones. 
Defining the angles $\theta_{1,2,\dots n}$ for the compactified
dimensions with $\theta_i \in [0,2\pi]$, we write 
them as  $x^c_i= R\theta_i$ where we assumed for simplicity
one common radius $R$. The Laplace equation may be written as
follows
\ba
\vec\nabla^2 \Phi &=& -
\delta^3(\vec{x}-\vec{y})
\frac{1}{R^n} \delta^n(\vec\theta-\vec\theta_0)
\label{L}
\ea
where the $\delta$--functions on the right-hand
side (RHS) are given as usually by
 \ba
\delta^3(\vec{x}-\vec{y})&=& \frac{1}{(2\pi)^3}\int d^3k e^{\imath
\vec{k}\cdot (\vec{x}-\vec{y})}\nonumber\\
\delta^n(\vec{x}^c-\vec{x}^c_0)&=&\frac{1}{(2\pi R)^n}\sum_{\vec{m}}
e^{\imath\vec{m}\cdot (\vec{\theta}^c-\vec{\theta}^c_0)}\nonumber
\ea
and the sums extend form $-\infty$ to $\infty$ for all indices
$m_{1,2\dots n}$. Using the Fourier transform, one finds
 \ba
 \Phi(r,q)&=&
\frac{1}{(2\pi)^{n+3}} \frac{1}{R^n}\sum_{\vec{m}}\int
d^3k\left\{e^{\imath\vec{k}\cdot\vec{r}+\imath
\vec{m}\cdot\vec{q}} \int_0^{\infty}d\,s
e^{-s[k^2+(\frac{\vec{m}}R)^2]}\right\} \label{ft}
\ea
where for simplicity we have denoted $\vec{r}=\vec{x}-\vec{y}$ and
$\vec{q}= \vec{\theta}-\vec{\theta_0}$.
In the integrand of (\ref{ft}), the summation is taken  over  the
infinite tower of KK--excitations in all the additional space
dimensions, $\vec{m}=(m_1,\dots m_n)$. It is easy now to perform
the integration with respect to $\vec{k}$. The result is
 \ba
\Phi(r,q) &=& \frac1{(4\pi)^{3/2}}\frac{1}{(2\pi R)^n}
\int_0^{\infty}d\,s s^{-3/2} e^{-\frac{r^2}{4 s}}
\sum_{\vec{m}}e^{\imath\vec{m}\cdot \vec{q}-s(\frac{\vec{m}}R)^2}
\label{V1}
\ea
 In the above summations,  $\vec{m}^2= m_1^2+\cdots + m_n^2$
 and $\vec{m}\cdot \vec{q} = m_1 q_1+\cdots m_n q_n$ is the inner
 product over  the $n-$dimensional compactified space.
 The  above result can also be written
in terms of  a product of $theta$ functions as follows
\ba
\Phi(r,q) &=& \frac{1}{(4\pi)^{3/2}}\frac{1}{(2\pi R)^n}
\int_0^{\infty}d\,s s^{-3/2} e^{-\frac{\vec{r}^2}{4
s}}\prod_{j=1}^n \theta_3\left(\frac{q_j}{2\pi},
\imath\frac{s}{\pi R^2}\right)
 \ea
where,
\ba \theta_3(\nu,\tau)
&=&\sum_{n=-\infty}^{\infty}p^{\frac{n^2}2}z^n
\ea
with $p= e^{2\imath\pi\tau}$ and $z= e^{2\imath\pi \nu}$.
Performing the integral we obtain
 \ba
\Phi(r,q)&=& \frac1{4\pi}\frac{1}{(2\pi R)^n}\frac1r \left\{
             1+2 \sum_{\vec{m}}^{\infty}
             e^{- |\vec{m}|\frac rR} \cos(\vec{m}\cdot\vec{q})\right\}
\label{Vnr}
\ea
 In the
particular case of one extra dimension, $n=1$, we may obtain an
exact result of the above integral.  We first perform the
integration in (\ref{V1}) to obtain
 \ba
\Phi(r,q)&=& \frac1{8\pi^2}\frac{1}{R}\frac1r \left\{
             1+2 \sum_{m=1}^{\infty}
             e^{- m\frac rR} \cos(m{q})\right\}
\label{VrgR}
 \ea
Performing the sum in this formula one gets the final expression
for $n=1$. Suppressing an overall numerical factor, we have the
following form for the potential
 \ba
  V_{n=1}(r) &\propto&\frac{1}{M^3_{P_5}}  \left(1+
2\frac{  e^{\frac{r}{R}}  \cos{q}-1}{ e^{2\frac{r}{R}}-  2
e^{\frac{r}{R}}\cos{q}+1 } \right)\,\frac{1}{R r} \label{ne1}
\ea
The dependence on the distance $r$ in this formula is exact and
valid for any value of $r$. For fixed $r$, its maximum value is
obtained when $q=0$, while for fixed $q$ the maxima are along the
path determined by the equation $r = R\log(1\pm \sin q)$. The
resulting potential as a function of $r$ and $q=\theta-\theta_0$
is plotted in figure 1.

In order to compare with the approximate formulae of the potential
given in the introduction,  we wish now to take the limit $q=0$ in
(\ref{ne1}) which gives
\ba
V&=&\frac{1}{M^3_{P_5}}
\frac{e^{\frac{r}{R}}+1}{e^{\frac{r}{R}}-1} \frac{1}{R r}
\label{exact}
\ea
The formula for $r\ll R$ becomes
 \ba
V_{n=1}\sim \frac{1}{M^3_{P_5}}\frac{2}{r^2} \left(1+ \frac 1{12}
\frac{r^2}{R^2}\right) \label{pa1}
\ea
This formula which is valid for  small $r$, differs by  a factor
of 2 compared to the approximation (\ref{ap1}). For $r\gg R$ we
obtain an exponential correction of the form
\ba
V_{n=1}&\sim &\frac{1}{M^3_{P_5}} \frac 1{R r} \left(1+2 e^{-\frac
rR}\right)\nonumber\\
       &\sim&\frac{M_C}{M^3_{P_5}} \frac 1{r} \left(1+2
e^{-{M_C}r}\right)\label{na1}
\ea
which is a Yukawa type correction valid for large distances
compared to the compactification radius.  The  approximation used,
gives us the chance to compare directly the above formula with the
usual parametrization of the long-range forces of gravitational
strength in the literature\cite{FT,cook}
\ba
V(r)\propto \frac 1r (1+\alpha e^{-\frac{r}{\lambda}})
 \label{par}.
\ea
Comparing the two formulae, we have a definite prediction for the
strength $\alpha$ of Yukawa type gravitational corrections in the case of
one extra compact dimension which is $\alpha =2$. Using the $\alpha$ -
$\lambda$ plot of\cite{lcp} which gives the experimentally
determined region, we conclude that the allowed radius has an
upper bound of the order $\lambda\equiv  R \sim 1 mm$.

Next, let us return to the approximate formulae in (\ref{ap1},
\ref{ap2}) which can be written in a single expression as,
 \ba
V\sim \frac{1}{M^3_{P_5}}\left(\frac 1{r^2}\theta(r-R) +\frac
1{r R} \theta(R-r)\right)\label{approx}.
\ea
 The formula (\ref{approx}) is plotted in figure 2 versus the
exact expression (\ref{exact}). The plot shows that  the two
expressions coincide only for $r\gg R$. For distances $r\sim R$
and $r<R$ there exist significant deviations which might lead to
interesting corrections in calculating various effects in physical
processes.


 {F}or more than one compact dimensions ($n>1$), we will work out
 approximated forms of the  potential.  As already stated, the
 approximations are straightforward in the case  where the radii of
 the extra compactified dimensions are either
 very  big or enormously small  compared to the distance that the
 potential is estimated, being those obtained from the
 Gauss' law in the `spherically' symmetric case. At relatively
large distances, $r>R$, we may also keep the first two terms of
the series expansion in (\ref{V1}), to obtain the result
\ba
\Phi(r,R)&=&\frac{1}{4\pi}\,\frac 1r \frac{1}{(2\pi R)^n
}\left(1+2 n e^{-r/R}\right)
\ea
which is a straightforward generalization of the approximation
(\ref{na1}) for arbitrary $n$. The other interesting case,
  which may have particular importance for the
experimental  verification of strongly coupled gravity at the TeV
scale, is when  the distance is comparable with the
compactification radius.

When the experimental measurement is taken in distances smaller
than the compactification radius of the extra dimensions $r \le
R$, the behavior of the infinite sum is not manifest since an
infinite number of terms may contribute. Then, the most effective
tool to extract the asymptotic behaviour of the potential in the
transition region where $R$ becomes effectively large, is the
Jacobi's transformation of theta functions\cite{book}
\ba
1+ 2\sum_{m=1}^{\infty} e^{-m^2\ell^2} \cos(2\pi m \ell z)&=&
\frac{\sqrt{\pi}}{\ell}e^{-\pi^2z^2}\left(1+2\sum_{m=1}^{\infty}
e^{-m^2\pi^2/\ell^2}\cosh\frac{2\pi^2m z}{\ell}\right).
\label{jth}
\ea
Substitution of the above formula in (\ref{V1}) gives
\ba
\Phi(r,q)&=& \frac{1}{(2\sqrt{\pi})^{n+3}}
\int_0^{\infty}
d\,s s^{-\frac{n+3}2}e^{-r^2/4s}\nonumber\\
 &\times&\prod_{j=1}^ne^{-q_j^2R^2/4 s}
 \left(1+2 \sum_{m_j}^{\infty}e^{-(m_j\pi R)^2/ s}
 \cosh\frac{m_jq_j\pi R^2}{s}
\right)
\label{VRgr}
\ea
Now, for  $R>r$  the exponentials in the sum converge rapidly and
a certain number of terms in the product may give a good
approximation.

We are interested in the case of two extra dimensions. 
Taking  the case
of zero angles, i.e. $q_j=\theta_j-\theta_{0j} = 0$ for all $j$'s 
and $n=2$ we can split (\ref{VRgr})  into three integrals which can
be evaluated. The results are,
\ba
I_0&=& \frac{1}{8\pi^2}\frac{1}{r^3}\label{i0}\\
I_1&=&\frac{4}{8\pi^2}\sum_{k}\frac{1}{(r^2+4\pi^2 k^2 R^2)^{3/2}}
\label{i1}\\
I_2&=&\frac{4}{8\pi^2}\sum_{k=1}\sum_{\ell=1}
\frac{1}{(r^2+4\pi^2 (k^2+\ell^2) R^2)^{3/2}}\label{i3}
\ea
We note that the number 4 multiplying the corrections
is the product of the factor 2 in front of the sum in
the integral (\ref{VRgr}) times the number of
dimensions $n=2$.
Defining the parameter  $\rho={r}/{(2\pi R)}$, for $\rho<1$
we may expand to obtain
\ba
I_1&\approx&\frac{1}{8\pi^2}\frac{4}{(2\pi R)^3}
         \left(\zeta(3)-\frac 32 \zeta(5)\rho^2\right)
\\
I_2&\approx&\frac{1}{8\pi^2}\frac{4}{(2\pi R)^3}
         \left(\zeta_2(3)-\frac 32 \zeta_2(5)\rho^2\right)
\ea
where in the above expressions $\zeta(\ell)$ is the Riemann zeta function
and we have introduced  the notation
$\zeta_2(\ell)=\sum_{k,m}(k^2+m^2)^{-\ell}$. Thus, we obtain
an approximation for the corrections
\ba
V(r)&\sim&\frac{1}{r^3}+4\frac{2.24}{(2\pi R)^3}
\ea
An estimation of the correction terms may also be given
in the limiting case $\rho\ra  1$. Putting $\rho=1$ and performing the
sums we obtain
\ba
V(r)&\sim&\frac{1}{r^3}+4\frac{1.32}{(2\pi R)^3}
\ea
The  general result
for $n=2$ can
be written as a double convergent sum as follows
\ba
\Phi_{n=2}&=&\frac 1{8\pi^2}\frac{1}{r^3} \left[1+ 4 {\rho}^3
\sum_{k=0}^{\infty}\sum_{l=1}^{\infty}
\frac{1}{\left(\rho^2+(k^2+l^2)\right)^{\frac 32}
}\right]\label{pa2}
\ea
The double sum
takes also into account  the degeneracy of  a
particular KK-contribution.
Clearly, the sum of the two integers $k^2+l^2= N^2$ which appears
in the denominator is related to the degeneracy.

 The generalization of the above result to higher dimensions is
straightforward. Here, we have restricted in examining the
 corrections to the Newtonian gravity due to the possible existence
of $n=1$ or $n=2$ extra space-time dimensions. We have succeeded
to obtain useful exact forms of the potential for the case
$n=1$. In the cases where $n>1$ the long range corrections can be
approximated by a Yukawa type interaction, and the potential is
written
$$V_{r>R}\sim \frac{1}{R^n r} (1+2 n e^{-r/R})$$
 where $r$ is the distance
and $R$ a common compactification radius of the $n$ extra
dimensions.  When the compact radius is effectively large, the
modifications are expressed as powers of the ratio $r/R$,
$$V_{r<R}\sim \frac{1}{ r^{n+1}}
\left(1+ 2 n c_n\left(\frac{r}{R}\right)^{n+1}\right)$$ where $c_n$ is a
calculable coefficient which for cases $n=1,2$ is given by the
expressions (\ref{pa1}) and (\ref{pa2}) in this work.

{\tt As this work was being written, we received \cite{ks} where a
similar analysis is presented.}

\vspace*{3cm}

\begin{center}
{\bf   FIGURE CAPTIONS }
\end{center}

{\bf FIGURE 1}: The potential $V(r)$ for one extra dimension
as a function of the distance-compactification ratio $r/R$
and the angle $q=\theta-\theta_0$.

{\bf FIGURE 2}: Comparison of the exact (upper curve) and approximate
forms  (lower curve) in the $n=1$ potential.
\newpage
\begin{figure}
\epsfig{figure=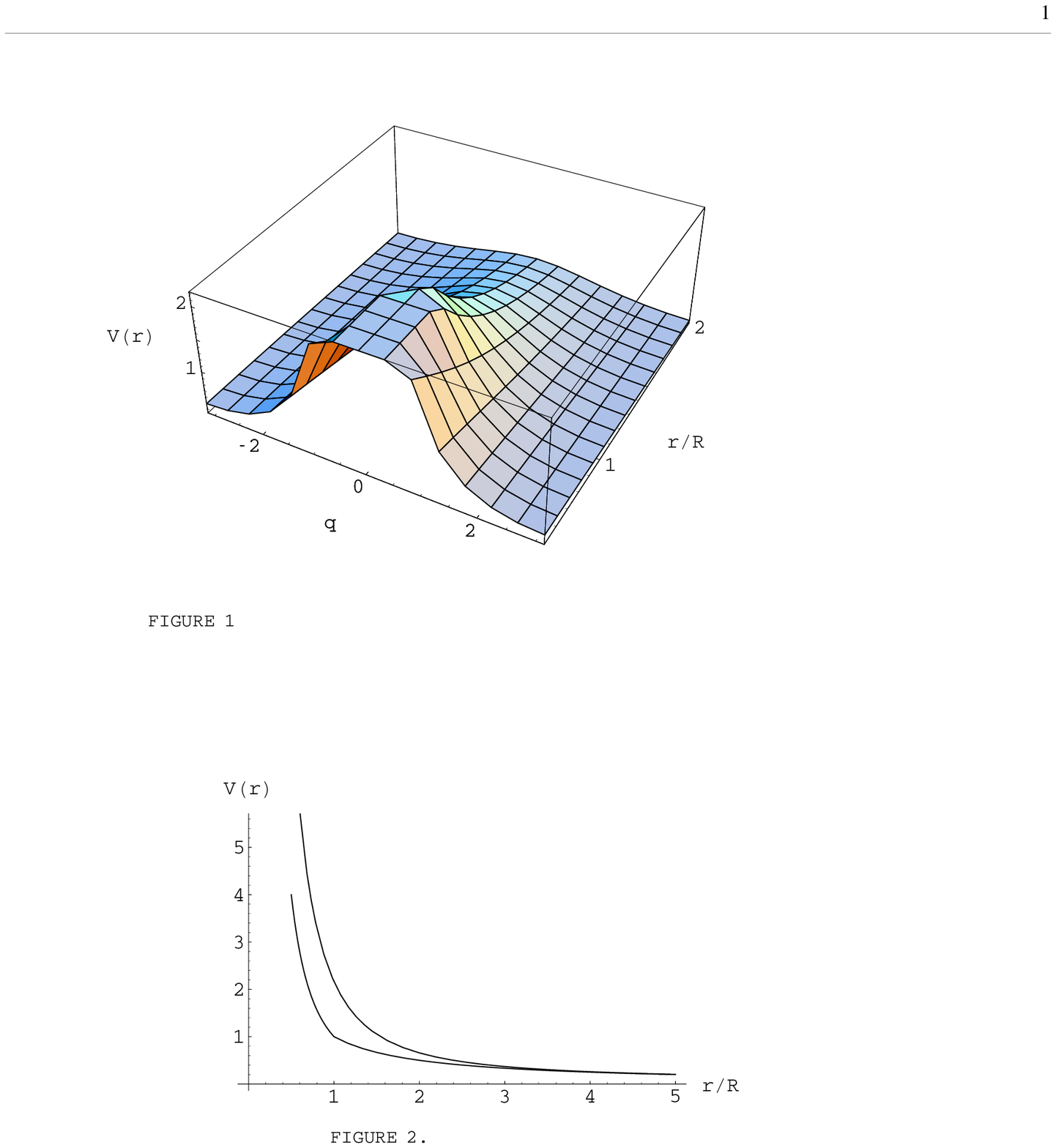}
\end{figure}

\end{document}